# Direct observation of locally modified excitonic effect within a moiré unit cell in twisted bilayer graphene

*Ming Liu[1], *Ryosuke Senga[2], Masanori Koshino[2], Yung-Chang Lin[2] and *Kazu Suenaga[1]*

[1]The Institute of Scientific and Industrial Research (SANKEN), Osaka University, Mihogaoka 8-1, Ibaraki, Osaka 567-0047, Japan

[2]Nanomaterials Research Institute, National Institute of Advanced Industrial Science and Technology (AIST), Higashi 1-1-1, Tsukuba 305-8565, Japan

**ABSTRACT**

Bilayer graphene, forming moiré superlattices, possesses distinct electronic and optical properties derived from the hybridization of energy band and the emergence of van Hove singularities depending on its twist angle. Extensive research has been conducted on the *global* characteristics of moiré superlattice induced by long-range periodicity. However, limited attention has been given to the *local* properties within a moiré unit cell, which undoubtedly differ due to the variations in three-dimensional atomic arrangement. Here we demonstrate the highly localized excitations of carbon 1s electrons to unoccupied van Hove singularities in a twisted bilayer graphene using an electron energy loss spectroscopy based on a monochromated transmission electron microscope. The core-level excitations associated with the van Hove singularities show a systematic twist angle dependence which is analogous to the optical excitations. Furthermore, local variations in those core-level van Hove singularity peaks within a moiré unit cell have been corroborated for the first time, which can originate from core-exciton lifetimes and band modifications influenced by the local stacking geometry.

**INTRODUCTION**

Moiré superlattices, which are created by stacking multiple layers of 2D materials on top of each other with a rotational mismatch, possess unique physical properties that cannot be found in individual 2D materials[1-6]. The moiré patterns provide a periodic potential landscape that can trap electrons and form specific band structures, including the energy band hybridization at the band crossing points and resulting van Hove singularities corresponding to the twist angles[7-9]. These tunable band structures have the potential to realize unique applications such as topological Hofstadter's butterfly[10, 11] and superconductivity[12, 13], with some of the magic twisting angles in bilayer graphene. The moiré superlattices also drastically modify optical properties to present distinct features such as moiré excitons in hetero/homo bilayer transition metal dichalcogenide (TMDC)[2-4, 14] and nonlinear excitonic transitions between the van Hove singularities in bilayer graphene[15].

The emergence of van Hove singularity in moiré superlattices, which strongly governs their electronic and optical properties, has been investigated through a range of techniques including optical spectroscopy[15-17], scanning tunneling spectroscopy (STS)[5, 6, 18-20] and electron energy loss spectroscopy (EELS)[21-23]. Optical spectroscopy has directly measured the optical resonances and demonstrated the strong excitonic effects in excitations from the van Hove singularities in the valence band to those in the conduction band at saddle points near the K point as well as rather higher energy excitations near the M point in bilayer graphene[15, 16]. STS has probed the surface charge density reflecting the moiré patterns and measured the transport gap between the van Hove singularities at the saddle points generated by interlayer coupling[13-15]. Although EELS can measure the optical gaps and excitonic interband transitions by valence-loss spectroscopy using an atomically sharp probe[21-23], the signal delocalization for those low-energy excitations caused by long-range Coulomb interaction has prevented atomically localized measurements[24]. Consequently, previous studies on moiré superlattices have mainly focused on the global electronic structures and optical properties and provided their twist angle dependences. The local properties that should differ among the various stacking geometries within a unit cell of moiré superlattice have not been highlighted.

In contrast, core-level EELS, which is attributed to localized core-electron excitations, has provided valuable insights into the chemical state of individual atoms[25]. Moreover, the recent advancements in TEM monochromators have significantly enhanced the energy resolution of EELS, facilitating the in-detail investigations of unoccupied states such as the van Hove singularities in the conduction band of carbon nanotubes[26].

In this study, we first investigate the twist angle dependence in excitations of valence and core electrons to the unoccupied states in bilayer graphene using high-energy resolution EELS with a monochromated electron source. The signature of the van Hove singularities resulting from the interlayer coupling is unambiguously identified by both valence-loss and core-loss spectroscopy. Then, by using an atomically sharp probe we clarify the locally modified excitonic features in core-electron excitations to the van Hove singularities in the conduction band within a single moiré unit cell taking advantage of the small signal delocalization in core-loss spectroscopy. Such locally probed electronic structures can lead to a deeper understanding of the underlying physics of moiré superlattices and open up new avenues for engineering the electronic and optical properties of van der Waals 2D materials for future applications.

**RESULTS**

Figure 1a shows an atomic model of a moiré superlattice induced by a twisted bilayer graphene with a twist angle of 9.8°. The moiré unit cell indicated by the dark blue line represents a visually periodic pattern, distinct from a rigorous unit cell that corresponds to the commensurability of the lattice period. When the two layers of graphene are stacked and coupled, the interlayer interaction perturbs the bands, leading to the hybridization of their energy band structures[22]. The Brillouin zone of the twisted bilayer graphene consists of two hexagonal Brillouin zones from different layers of graphene stacked at the twisted angle. The original K points of each graphene layer are denoted as $K_1$ and $K_2$, while the intersections of the hexagons are denoted as A and B, representing the regions where interlayer interactions and energy band overlap occur (Fig. 1b). The strong interlayer coupling results in energy band hybridization and the splitting of energy flat bands at the saddle points, giving rise to a minigap and new van Hove singularities[7].

In this study, we simultaneously acquired scanning transmission electron microscopy (STEM) images and EELS spectra using a transmission electron microscope (TEM) equipped with a double Wien filter type monochromator, along with STEM and TEM DELTA-type aberration correctors[27]. The energy resolution for low-loss spectroscopy was set at 35 meV, while 170 meV for core-loss spectroscopy (Supplementary Fig. 1). The electron beam geometry is provided in Fig. 1c and further detailed experimental conditions are presented in Method section.

The valence loss spectrum obtained from a twisted bilayer graphene with a twist angle of 9.8° reveals an additional sharp peak at approximately 1.8 eV, which is absent in the spectrum from a single-layer graphene (Fig. 1d). This peak corresponds to the excitation between the van Hove singularities in the valence and conduction bands. Additionally, the core-level EEL spectrum at the carbon K-edge from the same sample exhibits multiple peaks near the $\pi^*$ response, which correspond to the van Hove singularities in the conduction band and cannot be observed in the spectrum from a single layer graphene (Fig. 1e). Note that all spectra in this study were obtained from a nearly impurity-free region over several hundred nanometers at a consistent twisting angle, as shown in Supplementary Fig. 2i, to avoid any additional excitations associated with anti-bonding states of impurities such as amorphous carbon on the surface.

The schematic band structure of a twisted bilayer graphene is depicted in Figure 2a. The hybridization of energy bands occurs at A and B in the Brillouin zone, which serve as the saddle points of the energy band. An energy state corresponding to the van Hove singularity is denoted by $E_{xi}$ or $E_{xi}^*$, where x represents A, B, and M, denoting the saddle points. The index $i$ takes values 1 or 2, distinguishing between the bands that split at points A and B, while * indicates an anti-bonding state. An excitation from $E_{xi}$ to $E_{xj}^*$ is denoted as $E_{xij}$. A pair of van Hove singularities emerges in both the valence band ($E_{A1}$, $E_{A2}$) and the conduction band ($E_{A1}^*$, $E_{A2}^*$) at the saddle points A and B ($E_{B1}$, $E_{B2}$, $E_{B1}^*$,

$E_{B2}^*$) based of the previous theoretical studies[7, 28, 29]. Additionally, $E_M$ and $E_M^*$ represent intrinsic van Hove singularities at the local maximum of initial M point for each graphene layer, induced by the symmetry breaking of electrons and holes[15]. In far-field optical measurements, excitations are permitted from $E_{A1}$ ($E_{A2}$) to $E_{A2}^*$ ($E_{A1}^*$) at A and from $E_{B1}$ ($E_{B2}$) to $E_{B2}^*$ ($E_{B1}^*$) at B[7] (Fig. 2b). In the bright-field EELS, where dipole scattering is the primary contribution, all optically allowed excitations can be detected[30]. Therefore, the $E_A$ peak in Fig. 2c is mainly contributed by $E_{A12}$ ($E_{A1} \rightarrow E_{A2}^*$) and $E_{A21}$ ($E_{A2} \rightarrow E_{A1}^*$), which have nearly identical energies. It should be noted that the incident electron beam includes multiple momentum transfers ($0<q<\alpha$), thereby enabling the excitation of optically forbidden transitions, even though their inelastic scattering cross-section is considerably smaller than that of optically allowed transitions[31]. The minigap excitations at $K_1$ and $K_2$ are hidden in the tail of the quasi-elastic peak and difficult to resolve in our experimental condition. On the other hand, core-loss spectroscopy can detect excitations from 1s electrons to all unoccupied van Hove singularities as discussed later.

To investigate the twist angle dependence of optical excitations, we prepared twisted bilayer graphene samples with different twist angles by sequentially transferring two layers of graphene onto a TEM grid. The twist angles were confirmed through high-resolution TEM images and the corresponding fast Fourier transform (FFT) patterns as shown in Supplementary Figure 2. The $E_A$ peak including $E_{A12}$ and $E_{A21}$ shows a systematic blueshift as the twist angle increases (Fig. 2c), which is consistent with previous optical absorption and EELS studies[16, 22]. Both $E_B$ ($E_{B12}$ and $E_{B21}$) and $E_M$ excitations ($E_M \rightarrow E_M^*$) occur near the B point but are difficult to distinguish at the twist angles less than 20°. Only when the twist angle exceeds 20°, $E_B$ starts to separate form $E_M$, showing a redshift as the twist angle increases.

Figure 3 presents the twist angle dependence of the fine structure of carbon K-edge for bilayer graphene. Each spectrum in Fig. 3 is an average of more than 10 spectra from different positions for each twisting angle. The fine structure of the π* region obtained from the twist angle of 9.8° is shown with its fitting curves in Fig. 3a. The fine structure includes four components of C1s excitations to the van Hove singularities at A ($E_{A1}^*$, $E_{A2}^*$) and B ($E_{B1}^*$, $E_{B2}^*$) in the conduction band as well as the saddle points ($E_g^*$) at $K_1$ and $K_2$ which is the upper state of the mini gap (Fig. 3b). In addition to these five components, namely P1-P5, the overall π* resonance, in which the peak position and width are extracted from a single layer of graphene, and the excitations to the continuous state on the right side of the π* region are involved in the fittings. Note that the C1s excitation to the local maximum at the initial M point ($E_M^*$) is too close to the upper state of $E_B^*$ and thus is not distinguished as an isolated component in the fit. All the other spectra with different twist angles in Fig. 3c were fitted in the same way.

The energy positions of the van Hove singularity peaks extracted from the fit for each twist angle are shown in Supplementary Fig. 3. The peaks at the lowest energy (P1) assigned as the excitation to the saddle points at $K_1$ and $K_2$ are almost constant for all the twist angles since their band crossing points only shift near the Fermi level as the twist angle changes. The two peaks derived from $E_A^*$ (P2 and P3) increase while the other two derived from $E_B^*$ (P4 and P5) decrease as the twist angle increases. The energy level of upper $E_A^*$ ($E_{A2}^*$) and lower $E_B^*$ ($E_{B1}^*$) start to overlap at the twist angle of 25.7°. In this case, the peak at the highest energy (P5) in the 25.7° twist bilayer graphene can be assigned as the excitation to the local maximum at the M point ($E_M^*$), which is manifested by the degeneracy breaking of the $E_B^*$ and $E_M^*$. In the twist angles below 20° where $E_B^*$ and $E_M^*$ are close, the energy level of $E_B^*$ near the flat band should be almost constant. However, the peaks for $E_B^*$ at 4.8° and 6.4° shows a downshift. The twisted bilayer graphene with the twist angle close to 4°, which is often quoted to the threshold for the rigid atomic model[20], may start to be affected by local strains.

Since the signal delocalization of core-loss at carbon K-edge at 60kV is estimated as approximately 0.2 nm[24, 25], one can probe the local electronic structure even in a single unit cell of moiré superlattice. Figure 4a shows a STEM-ADF image of a moiré superlattice of the twist angle of 9.8°. The length of the unit cell of the moiré superlattice, shown as a rhombus, is about 1.5 nm. The simulated STEM image in Fig. 4a was constructed with a probe size of 2 Å, but no out-of-plane displacement nor the relaxation of moiré were considered. Note that the contrast in each moiré unit cell can be slightly different depending on the commensurability under the pure rigid model. However the influence of such higher-order moiré patterns is extremely complicated and beyond the scope of this study. Figure 4b shows the spectra obtained from three characteristic regions showing different contrast (we call hereafter; bright, grey and dark) in the moiré superlattice in Fig. 4a. The center of the bright, grey and dark region has AA, AB/BA and their mixture stacking so-called bridge structures, respectively. The spectra in Fig. 4c were constructed by summing 10~20 spectra obtained at equivalent structures within the same contrast in an impurity-free region covering 200 nanometers. Note that each spectrum integrates a 3 × 3 pixel region where the individual pixel size is approximately 0.15 nm squares. The three spectra extracted from different regions in the moiré unit cell do show the significant differences in π* peak. The same line shape analysis in Fig. 3 was performed for each spectrum. The energy positions and line widths of the five components are plotted in Fig. 4c. The spectrum at the bright region has multiple peaks in π* region derived from the sharp van Hove singularity peaks at $E_A$* and $E_B$* (P2-P5). In the grey region, although the positions of the van Hove singularity peaks hardly shift, their linewidth broadens by 5-40 % compared to the bright region. Similarly, the dark region shows a comparable broadening of 10-50% in the van Hove singularity peaks, even though the peak positions are almost identical to those of the bright region.

This broadening of the van Hove singularity peaks implies a reduction in the lifetime of core excitons. Interestingly, the core excitons exhibit longer lifetime in the bright region, where the interlayer interaction is supposedly weaker due to the energetically unfavorable AA stacking. The variation of exciton lifetime induced by the global staking geometry has been observed in $MoS_2$/$WSe_2$ van der Waals (vdW) heterostructures[21]. In this case of TMDC heterostructures, aligned stacking structures where the twist angle is close to 0° or 60° showed a faster interlayer charge transfer rate, leading to a shorter lifetime of interlayer excitons. Consequently, the exciton peak in TMDC heterostructures becomes broad in aligned stacking structures. Similarly, in the case of bilayer graphene, the local variation of charge transfer rate[32] may contribute to the lifetime reduction of core excitons in AB stacking and bridge structures compared to AA stacking.

In contrast to the peak width, the position and relative intensity of the van Hove singularity peaks in bilayer graphene with the twist angle of > 4.8° (Fig. 4) do not significantly change within a moiré unit cell. This result suggests that the local band structure does not undergo significant changes within the moiré unit cell when the twist angle is sufficiently large for a rigid model to be applicable. For a twist angle of 6.4° in bilayer graphene as shown in Supplementary Fig. 4c,d, the linewidth of the van Hove singularity peaks broadens in AB and bridge regions, similar to the case of 9.8°, while the peak positions shift more by 200-300 meV. As the twist angle decreases to 4.8°, the local structure dependence of the fine structure in the π* region becomes more pronounced due to both peak shifts and broadening of each van Hove singularity peak (Supplementary Fig. 4e,f). In this smaller twist angle, the reconstruction starts to take place and provide the modified band structure due to the local strain[9, 20].

**CONCLUSION**

Our method successfully probes the atomically localized unoccupied states of bilayer graphene with an energy resolution of better than 170 meV, which is comparable to the standard X-ray absorption spectroscopy. This powerful tool enables the investigation of local electronic structures, electron-electron and electron-hole interactions, and thus provides the insights into the origins of optical and electronic properties of vdW 2D materials. Combining of further theoretical calculations including core-hole effects, which are beyond the scope of this study, has a potential to elucidate the fundamental physics in 2D materials.

**EXPERIMENTAL SECTION**

[Sample preparations]

The graphene used in the experiments was grown on Cu foil by plasma chemical vapor deposition (CVD), which consists of minimal number of defects[33]. Twisted bilayer graphene with different twist angles were obtained by transferring two layers of graphene separately to TEM grids with a well-developed clean transfer process[34, 35]. The samples were heated at 500 °C overnight in the TEM.

[STEM/TEM and EELS]

The high energy resolution EELS was performed using a TEM (JEOL Triple C#2) equipped with STEM and TEM DELTA-type spherical aberration correctors with a double Wien filter type monochromator. The accelerating voltage was 60 kV for all experiments. The probe current was 8 and 20 pA for low-loss and core-loss spectroscopy, respectively. The convergence semi-angle was 50 mrad for all STEM imaging and spectroscopy. The inner semi-angle of ADF detector and EELS collection semi-angle was set to approximately 55 and 60 mrad, respectively for low-loss spectroscopy while 55 and 80 mrad were used for core-loss spectroscopy. STEM images of moiré superlattices were acquired with a JEOL ADF detector optimized for middle- and low-angle ADF imaging. The STEM image simulation in Fig. 4 was performed by TEMPAS (Total Resolution LLC). EEL spectra were acquired using a Gatan GIF spectrometer optimized for low-acceleration voltages, where the electron convergent efficiency of the scintillator offers 68counts/electron at 60kV. The full-width at half-maximum of the energy resolution of the electron probe was 170 meV for core-loss spectroscopy and 35 meV for low-loss spectroscopy. In our core-loss spectroscopy, we chose the energy dispersion of 50 meV per channel, which is close to the energy resolution of the incident beam. This setup gives us an excellent balance between the energy resolution and signal to noise ratio over the wider energy range required for the carbon K-edge analysis. On the other hand, this configuration results in a rather smoothed tail of the zero-loss peak due to the point spread function of the CCD camera, making it more like a Gaussian shape. Therefore we eventually chose the fitting with Gaussian components which works best for our results. The core-loss and low-loss spectra were energetically calibrated using unsaturated zero-loss peaks obtained simultaneously. All twist angles were obtained from the high-resolution TEM images and their FFT images acquired with the same microscope in TEM mode using a Gatan Oneview camera optimized for low-acceleration voltage. All the TEM and EELS experiments were performed at 500 °C.

[Data analysis]

The background of the low-loss spectra shown in Fig. 1(d) and Fig.2(c) was subtracted with a power-law model. The core-loss spectra shown in Figures 3 and 4 were processed by background subtraction with a power-law model and then deconvolutions with the simultaneously acquired zero-loss peak as the Kernel spectrum using a maximum entropy method. The line shape analysis on the spectra in Figures 3, 4 and Extended Figure 3 was performed with five Gaussian components (P1-P5) and rather broad Gaussian components for the overall $\pi^*$ resonance and the energy-loss higher than 286.5 eV. This Gaussian assumption for the overall p* resonance delocalized in the momentum space has been often used and provided reasonable fitting results in the analysis of carbon K-edge in X-ray absorption spectroscopy[36, 37]. The peak position and width for the overall $\pi^*$ resonance were extracted from a single layer graphene and fixed for all other spectra in the process. The fitting process was performed through an iterative algorithm using the Levenberg-Marquardt method, within the framework of a commercial software package (OriginPro 2023). This algorithm iteratively proceeded until the reduction of the Chi-square statistic reached a value below $10^{-6}$. During the iterative procedure, the parameters associated with the position, intensity, and width of the two Gaussian components relating to the overall π* resonance and energy-loss exceeding 286.5 eV were held constant. Conversely, the parameters governing the van Hove singularity peaks (P1-P5) were allowed to vary without constraint, except for the number of components and the boundary condition concerning the peak positions (P1 < P2 < P3 < P4< P5). The uncertainties of the fitting parameters were described by the standard errors based on the Variance-Covariance matrix of the fitting parameters, represented by the error bars in Fig. 4d. The minimum residue was obtained by using Gaussian for all the components among the fits with Gaussian, Lorentzian and Voigt functions because of the Gaussian-like profile of the zero-loss peak due to the current limitation by the experimental condition described above.

**ASSOCIATED CONTENTS**

**Supporting information**

All supplementary figures.

**Author contributions**

RS and KS designed the project. ML and RS performed all TEM experiments. ML, RS and MK analysed the data. ML and YCL prepared the samples. ML, RS and KS co-wrote the paper. All authors commented on the manuscript.

**Notes**

The authors declare no competing financial interests.

**AUTHOR INFORMATION**

**Corresponding Author**

Ryosuke Senga (ryosuke-senga@aist.go.jp)

Kazu Suenaga (suenaga-kazu@sanken.osaka-u.ac.jp)

**ACKNOWLEDGEMENTS**

This work was supported by JST-CREST (JPMJCR20B1, JPMJCR20B5, JPMJCR1993), JST-PRESTO (JPMJPR2009), JSPS KAKENHI (19K04434, 21H05235, 22H05478, 23H00277, 23H01807) and ERC "MORE-TEM (951215)".

# REFERENCES

1. Lau, C. N.; Bockrath, M. W.; Mak, K. F.; Zhang, F., Reproducibility in the fabrication and physics of moiré materials. *Nature* **2022,** *602*, 41-50, DOI: 10.1038/s41586-021-04173-z.

2. Tran, K.; Moody, G.; Wu, F.; Lu, X.; Choi, J.; Kim, K.; Rai, A.; Sanchez, D. A.; Quan, J.; Singh, A.; Embley, J.; Zepeda, A.; Campbell, M.; Autry, T.; Taniguchi, T.; Watanabe, K.; Lu, N.; Banerjee, S. K.; Silverman, K. L.; Kim, S.; Tutuc, E.; Yang, L.; MacDonald, A. H.; Li, X., Evidence for moiré excitons in van der Waals heterostructures. *Nature* **2019,** *567*, 71-75, DOI: 10.1038/s41586-019-0975-z.

3. Seyler, K. L.; Rivera, P.; Yu, H.; Wilson, N. P.; Ray, E. L.; Mandrus, D. G.; Yan, J.; Yao, W.; Xu, X., Signatures of moiré-trapped valley excitons in MoSe2/WSe2 heterobilayers. *Nature* **2019,** *567*, 66-70, DOI: 10.1038/s41586-019-0957-1.

4. Jin, C.; Regan, E. C.; Yan, A.; Iqbal Bakti Utama, M.; Wang, D.; Zhao, S.; Qin, Y.; Yang, S.; Zheng, Z.; Shi, S.; Watanabe, K.; Taniguchi, T.; Tongay, S.; Zettl, A.; Wang, F., Observation of moiré excitons in WSe2/WS2 heterostructure superlattices. *Nature* **2019,** *567*, 76-80, DOI: 10.1038/s41586-019-0976-y.

5. Kerelsky, A.; McGilly, L. J.; Kennes, D. M.; Xian, L.; Yankowitz, M.; Chen, S.; Watanabe, K.; Taniguchi, T.; Hone, J.; Dean, C.; Rubio, A.; Pasupathy, A. N., Maximized electron interactions at the magic angle in twisted bilayer graphene. *Nature* **2019,** *572*, 95-100, DOI: 10.1038/s41586-019-1431-9.

6. Jiang, Y.; Lai, X.; Watanabe, K.; Taniguchi, T.; Haule, K.; Mao, J.; Andrei, E. Y., Charge order and broken rotational symmetry in magic-angle twisted bilayer graphene. *Nature* **2019,** *573*, 91-95, DOI: 10.1038/s41586-019-1460-4.

7.	Moon, P.; Koshino, M., Optical absorption in twisted bilayer graphene. *Physical Review B* **2013,** *87*, 205404, DOI: 10.1103/PhysRevB.87.205404.

8.	Bistritzer, R.; MacDonald, A. H., Moiré bands in twisted double-layer graphene. *Proc. Natl. Acad. Sci. U. S. A.* **2011,** *108*, 12233-12237, DOI: 10.1073/pnas.1108174108.

9.	Nam, N. N. T.; Koshino, M., Lattice relaxation and energy band modulation in twisted bilayer graphene. *Physical Review B* **2017,** *96*, 075311, DOI: 10.1103/PhysRevB.96.075311.

10.	Nemec, N.; Cuniberti, G., Hofstadter butterflies of bilayer graphene. *Physical Review B* **2007,** *75*, 201404(R), DOI: 10.1103/PhysRevB.75.201404.

11.	Bistritzer, R.; MacDonald, A. H., Moiré butterflies in twisted bilayer graphene. *Physical Review B* **2011,** *84*, 035440, DOI: 10.1103/PhysRevB.84.035440.

12.	Yankowitz, M.; Chen, S.; Polshyn, H.; Zhang, Y.; Watanabe, K.; Taniguchi, T.; Graf, D.; Young, A. F.; Dean, C. R., Tuning superconductivity in twisted bilayer graphene. *Science* **2019,** *363*, 1059-1064, DOI: 10.1126/science.aav1910.

13.	Oh, M.; Nuckolls, K. P.; Wong, D.; Lee, R. L.; Liu, X.; Watanabe, K.; Taniguchi, T.; Yazdani, A., Evidence for unconventional superconductivity in twisted bilayer graphene. *Nature* **2021,** *600*, 240-245, DOI: 10.1038/s41586-021-04121-x.

14.	Huang, D.; Choi, J.; Shih, C. K.; Li, X., Excitons in semiconductor moiré superlattices. *Nature Nanotechnology* **2022,** *17*, 227-238, DOI: 10.1038/s41565-021-01068-y.

15.	Chae, D. H.; Utikal, T.; Weisenburger, S.; Giessen, H.; Klitzing, K. V.; Lippitz, M.; Smet, J., Excitonic fano resonance in free-standing graphene. *Nano Letters* **2011,** *11*, 1379-1382, DOI: 10.1021/nl200040q.

16. Havener, R. W.; Liang, Y.; Brown, L.; Yang, L.; Park, J., Van Hove singularities and excitonic effects in the optical conductivity of twisted bilayer graphene. *Nano Letters* **2014,** *14*, 3353-3357, DOI: 10.1021/nl500823k.

17. Wang, T.; Liu, Q. F.; Caraiani, C.; Zhang, Y. P.; Wu, J.; Chan, W. L., Effect of Interlayer Coupling on Ultrafast Charge Transfer from Semiconducting Molecules to Mono- and Bilayer Graphene. *Physical Review Applied* **2015,** *4*, 014016, DOI: 10.1103/PhysRevApplied.4.014016.

18. Choi, Y.; Kemmer, J.; Peng, Y.; Thomson, A.; Arora, H.; Polski, R.; Zhang, Y.; Ren, H.; Alicea, J.; Refael, G.; von Oppen, F.; Watanabe, K.; Taniguchi, T.; Nadj-Perge, S., Electronic correlations in twisted bilayer graphene near the magic angle. *Nature Physics* **2019,** *15*, 1174-1180, DOI: 10.1038/s41567-019-0606-5.

19. Wong, D.; Wang, Y.; Jung, J.; Pezzini, S.; DaSilva, A. M.; Tsai, H. Z.; Jung, H. S.; Khajeh, R.; Kim, Y.; Lee, J.; Kahn, S.; Tollabimazraehno, S.; Rasool, H.; Watanabe, K.; Taniguchi, T.; Zettl, A.; Adam, S.; MacDonald, A. H.; Crommie, M. F., Local spectroscopy of moiré-induced electronic structure in gate-tunable twisted bilayer graphene. *Physical Review B* **2015,** *92*, 155409, DOI: 10.1103/PhysRevB.92.155409.

20. Yoo, H.; Engelke, R.; Carr, S.; Fang, S.; Zhang, K.; Cazeaux, P.; Sung, S. H.; Hovden, R.; Tsen, A. W.; Taniguchi, T.; Watanabe, K.; Yi, G. C.; Kim, M.; Luskin, M.; Tadmor, E. B.; Kaxiras, E.; Kim, P., Atomic and electronic reconstruction at the van der Waals interface in twisted bilayer graphene. *Nature Materials* **2019,** *18*, 448-453, DOI: 10.1038/s41563-019-0346-z.

21. Gogoi, P. K.; Lin, Y. C.; Senga, R.; Komsa, H. P.; Wong, S. L.; Chi, D.; Krasheninnikov, A. V.; Li, L. J.; Breese, M. B. H.; Pennycook, S. J.; Wee, A. T. S.; Suenaga, K., Layer Rotation-Angle-Dependent Excitonic Absorption in van der Waals

Heterostructures Revealed by Electron Energy Loss Spectroscopy. *ACS Nano* **2019**, *13*, 9541-9550, DOI: 10.1021/acsnano.9b04530.

22.     Lin, Y. C.; Motoyama, A.; Solis-Fernandez, P.; Matsumoto, R.; Ago, H.; Suenaga, K., Coupling and Decoupling of Bilayer Graphene Monitored by Electron Energy Loss Spectroscopy. *Nano Letters* **2021,** *21*, 10386-10391, DOI: 10.1021/acs.nanolett.1c03689.

23.     Susarla, S.; Naik, M. H.; Blach, D. D.; Zipfel, J.; Taniguchi, T.; Watanabe, K.; Huang, L.; Ramesh, R.; da Jornada, F. H.; Louie, S. G.; Ercius, P.; Raja, A., Hyperspectral imaging of exciton confinement within a moiré unit cell with a subnanometer electron probe. *Science* **2022,** *378*, 1235-1239, DOI: 10.1126/science.add9294.

24.     Egerton, R. F., Limits to the spatial, energy and momentum resolution of electron energy-loss spectroscopy. *Ultramicroscopy* **2007,** *107*, 575-586, DOI: 10.1016/j.ultramic.2006.11.005.

25.     Suenaga, K.; Koshino, M., Atom-by-atom spectroscopy at graphene edge. *Nature* **2010,** *468*, 1088-1090, DOI: 10.1038/nature09664.

26.     Senga, R.; Pichler, T.; Suenaga, K., Electron Spectroscopy of Single Quantum Objects To Directly Correlate the Local Structure to Their Electronic Transport and Optical Properties. *Nano Letters* **2016,** *16*, 3661-3667, DOI: 10.1021/acs.nanolett.6b00825.

27.     Hosokawa, F.; Sawada, H.; Kondo, Y.; Takayanagi, K.; Suenaga, K., Development of Cs and Cc correctors for transmission electron microscopy. *Microscopy (Oxf)* **2013,** *62*, 23-41, DOI: 10.1093/jmicro/dfs134.

28.	Wang, J.; Bo, W.; Ding, Y.; Wang, X.; Mu, X., Optical, optoelectronic, and photoelectric properties in moiré superlattices of twist bilayer graphene. *Materials Today Physics* **2020,** *14*, 100238, DOI: 10.1016/j.mtphys.2020.100238.

29.	Yang, L.; Deslippe, J.; Park, C. H.; Cohen, M. L.; Louie, S. G., Excitonic effects on the optical response of graphene and bilayer graphene. *Physical Revew Letters* **2009,** *103*, 186802, DOI: 10.1103/PhysRevLett.103.186802.

30.	Hong, J.; Senga, R.; Pichler, T.; Suenaga, K., Probing Exciton Dispersions of Freestanding Monolayer WSe2 by Momentum-Resolved Electron Energy-Loss Spectroscopy. *Physical Review Letters* **2020,** *124*, 087401, DOI: 10.1103/PhysRevLett.124.087401.

31.	Senga, R.; Pichler, T.; Yomogida, Y.; Tanaka, T.; Kataura, H.; Suenaga, K., Direct Proof of a Defect-Modulated Gap Transition in Semiconducting Nanotubes. *Nano Lett* **2018,** *18*, 3920-3925, DOI: 10.1021/acs.nanolett.8b01284.

32.	Yu, Y.; Zhang, K.; Parks, H.; Babar, M.; Carr, S.; Craig, I. M.; Van Winkle, M.; Lyssenko, A.; Taniguchi, T.; Watanabe, K.; Viswanathan, V.; Bediako, D. K., Tunable angle-dependent electrochemistry at twisted bilayer graphene with moiré flat bands. *Nature Chemistry* **2022,** *14*, 267-273, DOI: 10.1038/s41557-021-00865-1.

33.	Kato, R.; Minami, S.; Koga, Y.; Hasegawa, M., High growth rate chemical vapor deposition of graphene under low pressure by RF plasma assistance. *Carbon* **2016,** *96*, 1008-1013, DOI: 10.1016/j.carbon.2015.10.061.

34.	Lin, Y. C.; Jin, C.; Lee, J. C.; Jen, S. F.; Suenaga, K.; Chiu, P. W., Clean transfer of graphene for isolation and suspension. *ACS Nano* **2011,** *5*, 2362-2368, DOI: 10.1021/nn200105j.

35. Lin, Y. C.; Lu, C. C.; Yeh, C. H.; Jin, C.; Suenaga, K.; Chiu, P. W., Graphene annealing: how clean can it be? *Nano Lett* **2012,** *12*, 414-419, DOI: 10.1021/nl203733r.

36. Kramberger, C.; Rauf, H.; Shiozawa, H.; Knupfer, M.; Büchner, B.; Pichler, T.; Batchelor, D.; Kataura, H., Unraveling van Hove singularities in x-ray absorption response of single-wall carbon nanotubes. *Physical Review B* **2007,** *75*, DOI: 10.1103/PhysRevB.75.235437.

37. De Blauwe, K.; Mowbray, D. J.; Miyata, Y.; Ayala, P.; Shiozawa, H.; Rubio, A.; Hoffmann, P.; Kataura, H.; Pichler, T., Combined experimental and ab initio study of the electronic structure of narrow-diameter single-wall carbon nanotubes with predominant (6,4),(6,5) chirality. *Physical Review B* **2010,** *82*, 125444, DOI: 10.1103/PhysRevB.82.125444.

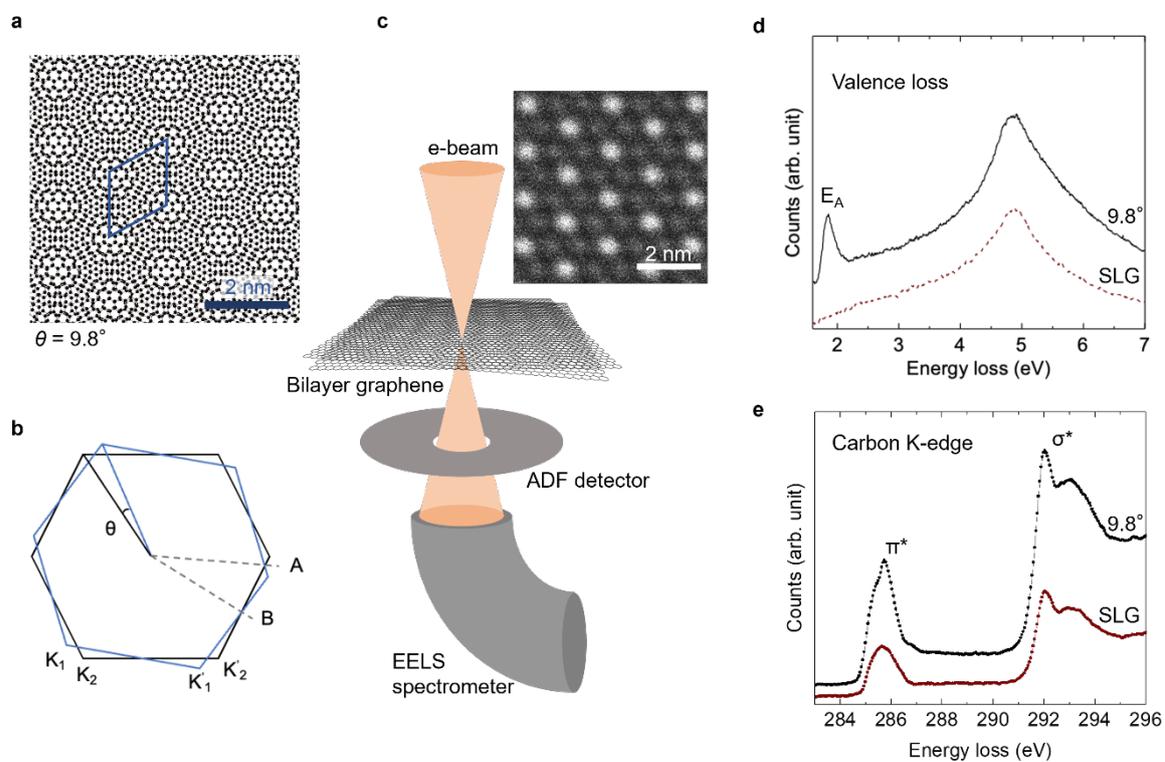

**Figure 1. Valence loss spectrum and Carbon *K*-edge spectrum from a moiré superlattice of a twisted bilayer graphene.** (a) Schematic of a moiré superlattice induced by the twisted neighboring two layers in a spiral graphene nanosheet in top view. (b) Schematic of the Brillouin zone of graphene overlapped with twist angle $\theta$. (c) Electron beam geometry of EELS condition and a STEM-ADF (annular dark-field) image of a twisted bilayer graphene with 9.8°. (d, e) Representative valence loss spectra and Carbon *K*-edge spectrum from bilayer graphene with a twist angle ($\theta$=9.8°), respectively. Valence- and core-loss spectra of a single layer graphene (SLG) are accompanied as a reference. The van Hove singularity peak ($E_A$) is labelled in (d) The $\pi^*$ and $\sigma^*$ regions are marked in (e).

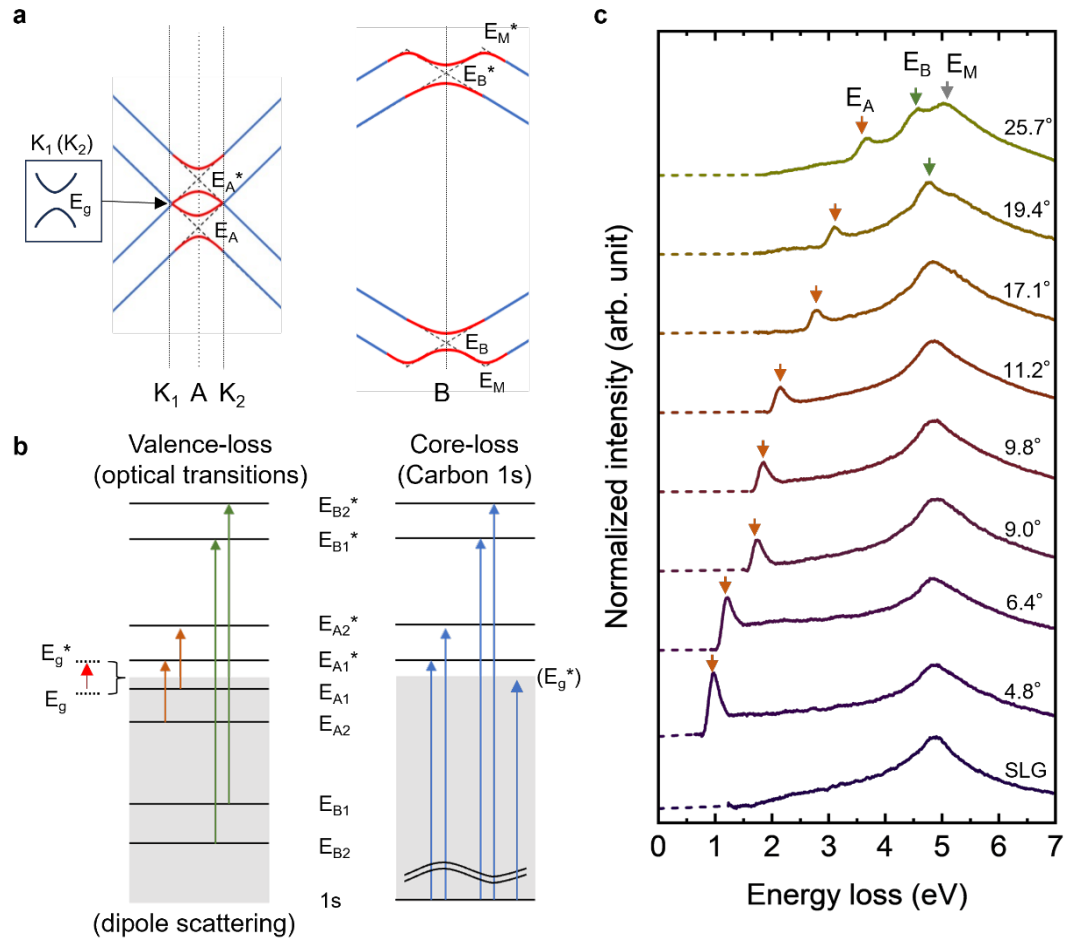

**Figure 2. Twist angle dependence of the valence and core excitations associated with the van Hove singularities.** (a) Schematic band structures of a twisted bilayer graphene in the zone of A and B in Fig. 1b based on the previous band theory[7]. (b) Comparison of excitations in valence- and core-loss (carbon 1s) spectroscopy with simplified schematic of possible excitations associated with the van Hove singularities. The possible transitions in valence-loss spectroscopy at dipole scattering in the long wavelength limit are basically equivalent to the optically allowed transitions reported in Ref. 16. (c) Twist angle

dependence of the valence-loss spectra from BLGs. $E_A$ ($E_B$) includes the excitations of $E_{A1}$ ($E_{B1}$) → $E_{A2}^*$ ($E_{B2}^*$) and $E_{A2}$ ($E_{B2}$) → $E_{A1}^*$ ($E_{B1}^*$). The spectrum from a single layer graphene is also presented as a reference. While the solid lines present the background subtracted spectra, the dotted lines correspond to the region where the background cannot be accurately subtracted. Note that we only discuss the relative peak positions here since it is hardly possible to obtain quantitative loss functions.

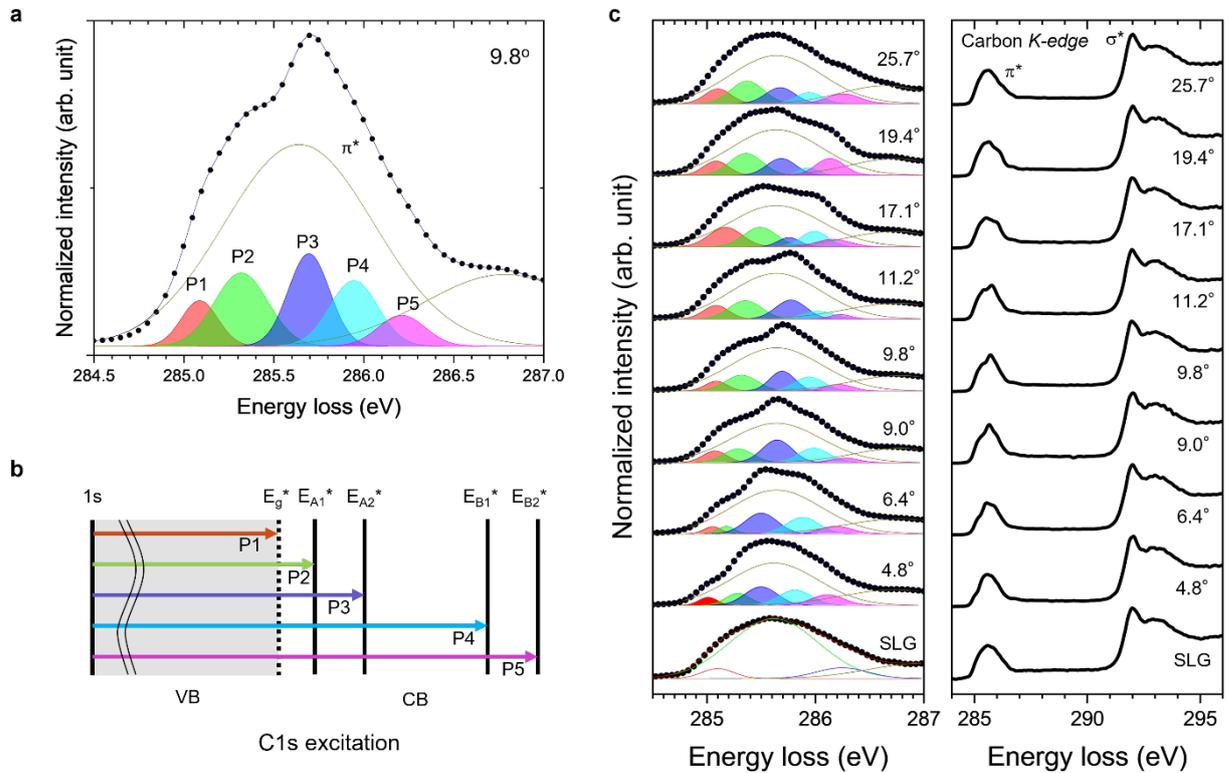

**Figure 3. Twist angle dependence of the fine structure of carbon K-edge.** (a) Fine structures of the π* region of the carbon K-edge obtained from a twist angle of 9.8°. A line shape analysis consisting of five components of the van Hove singularity peaks (P1 to P5) along with the overall π* resonance and the energy-loss for the end of the π* region is used to fit the observed fine structure. (b) Simplified schematic of the C1s excitations to the van Hove singularities in the conduction bands corresponding to peaks P1 to P5. (c) A series of the fine structures of the carbon K-edge (right panel) and magnified π* region (left panel) obtained from all measured twist angles with the results of the line

shape analysis using the same number of components as in (a). The one for a single layer graphene is shown as a reference.

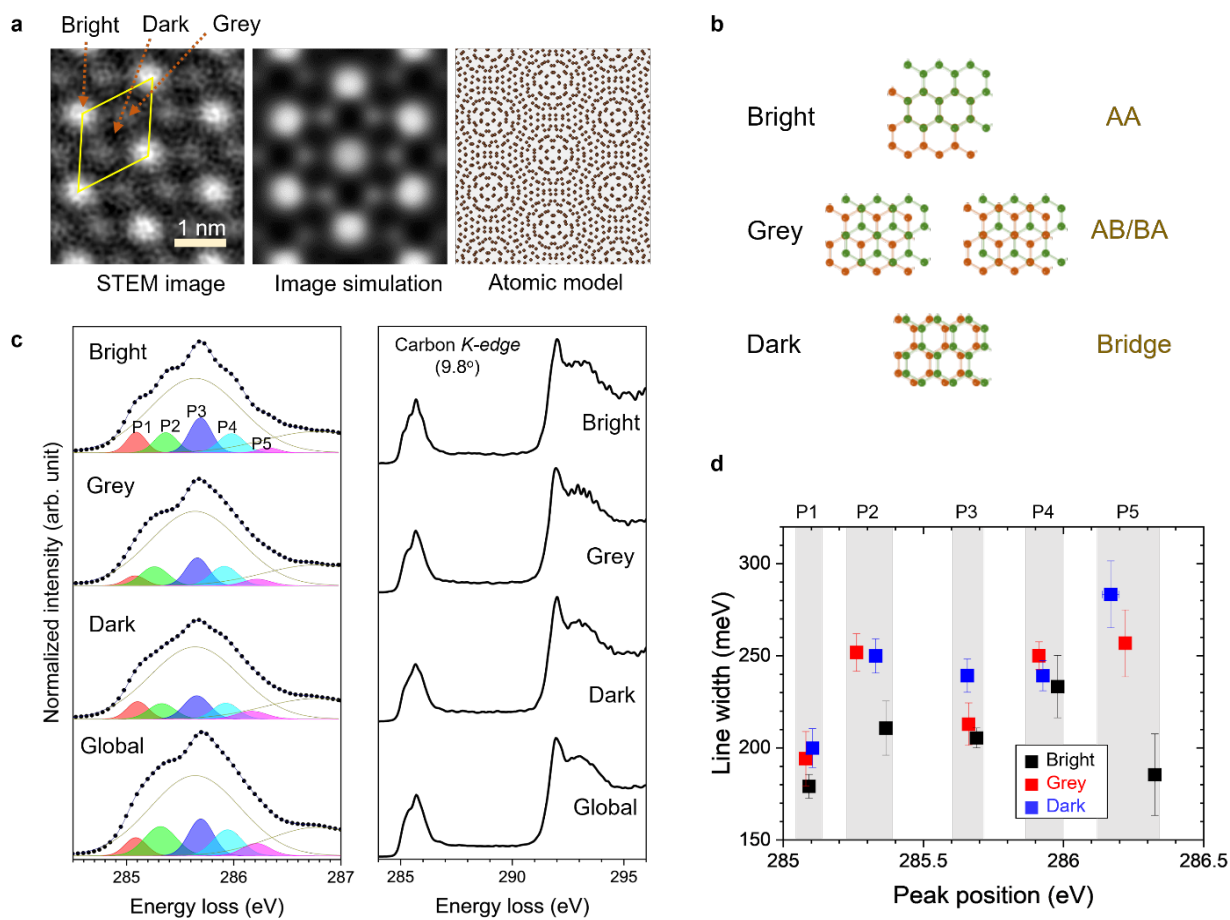

**Figure 4. Local structure dependence of carbon K-edge fine structure within a moiré unit cell.** (a) Drift-corrected STEM image of a moiré superlattice with the twist angle of 9.8° (left) and its corresponding simulated STEM image (middle) and atomic model (right). (b) Atomic structures of bright, grey, and dark regions in (a). (c) The fine structures of the carbon K-edge (right panel) and the magnified π* region (left panel) corresponding to the bright, grey and dark regions with the results of the line shape analysis as well as the global spectra integrated over a moiré unit cell. The peak position

and line width of peaks P1 to P5 for the bright, grey and dark regions in (c) are shown in (d) as black, red and blue dots, respectively with the error bars indicating the fitting errors.

SUPPORTING INFORMATION

# Direct observation of locally modified excitonic effect within a moiré unit cell in twisted bilayer graphene

*Ming Liu[1], \*Ryosuke Senga[2], Masanori Koshino[2], Yung-Chang Lin[2] and \*Kazu Suenaga[1]*

[1]The Institute of Scientific and Industrial Research (SANKEN), Osaka University, Mihogaoka 8-1, Ibaraki, Osaka 567-0047, Japan

[2]Nanomaterials Research Institute, National Institute of Advanced Industrial Science and Technology (AIST), Higashi 1-1-1, Tsukuba 305-8565, Japan

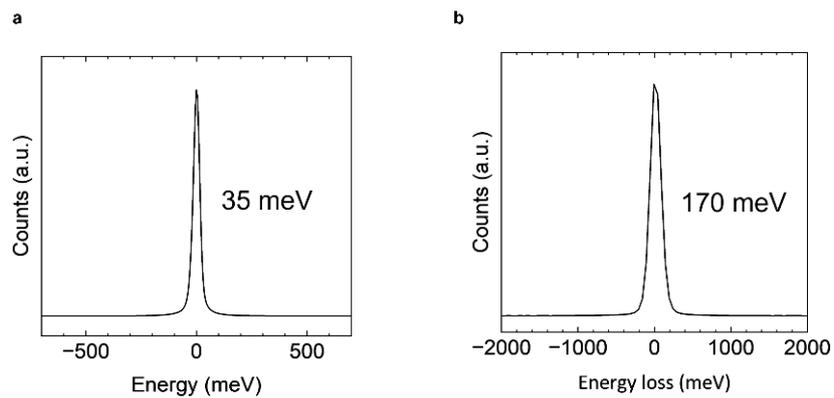

**Supplementary Figure 1. Energy resolution of low-loss and core-loss EELS experimental conditions. a**, Zero-loss peak from an empty spot measured with low-loss experimental conditions, FWHM=35 meV. **b**, Zero-loss peak from an empty spot measured with core-loss experimental conditions, FWHM=170 meV.

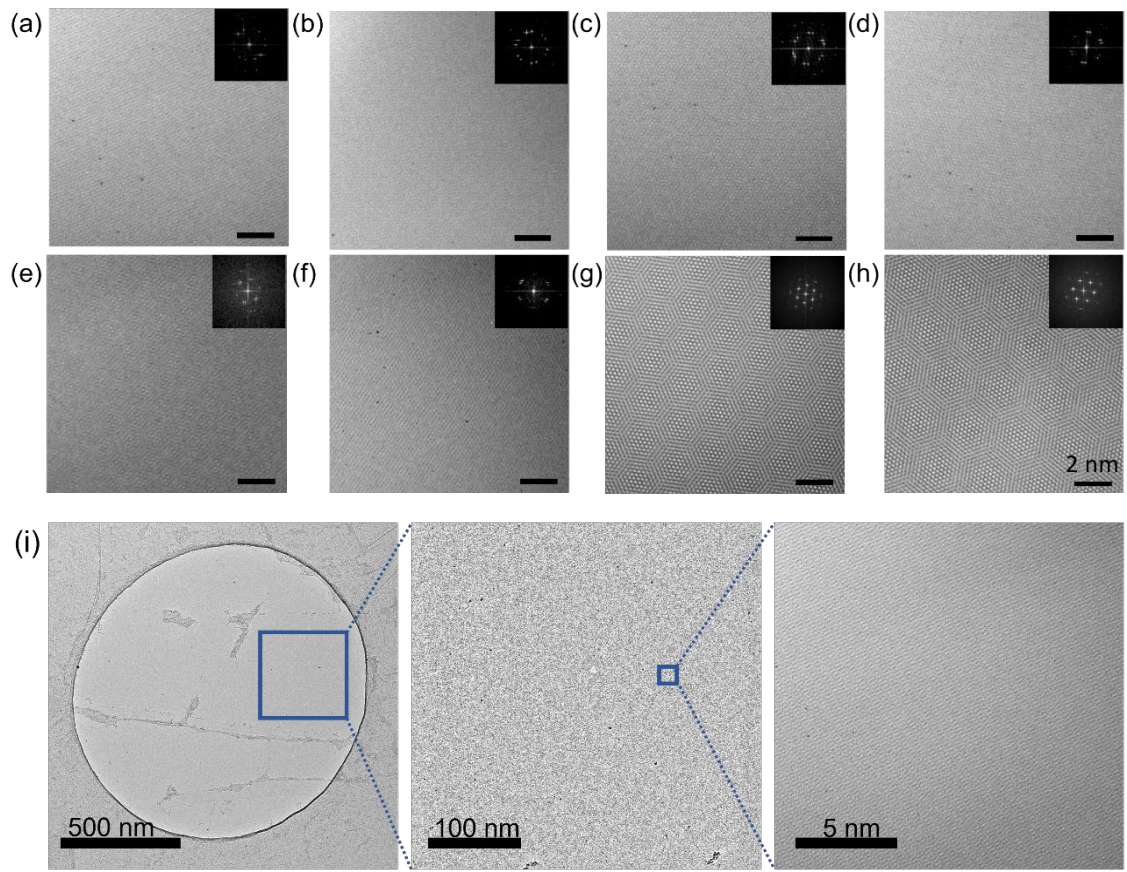

**Supplementary Figure 2. High resolution TEM images and FFT patterns (inset) of the examined twisted bilayer graphene. a-h**, Twisted bilayer graphene with twist angles of 25.7°, 19.4°, 17.1°, 11.2°, 9.8°, 9.0°, 6.4°, and 4.8°, respectively. **i**, Low-magnification TEM images for the bilayer graphene with twist angle of 25.7°.

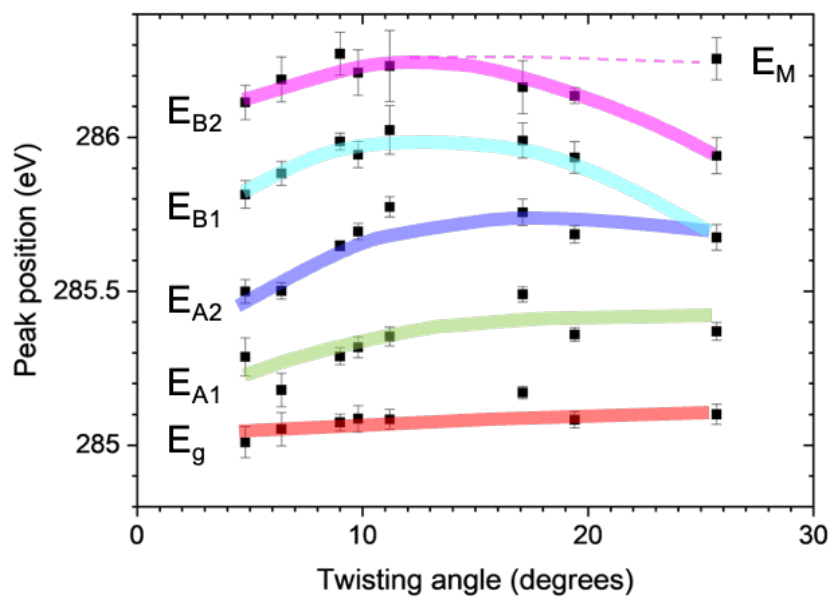

**Supplementary Figure 3. Twist angle dependence of the van Hove singularity peaks in the carbon K-edge.** The peak positions of each component extracted from the line shape analysis are plotted with the error bars indicating the fitting errors. The possible assignments of $E_g^*$, $E_{A1}^*$, $E_{A2}^*$, $E_{B1}^*$ and $E_{B2}^*$ are represented by the guidelines shown in red, green, blue, light bule and purple, respectively.

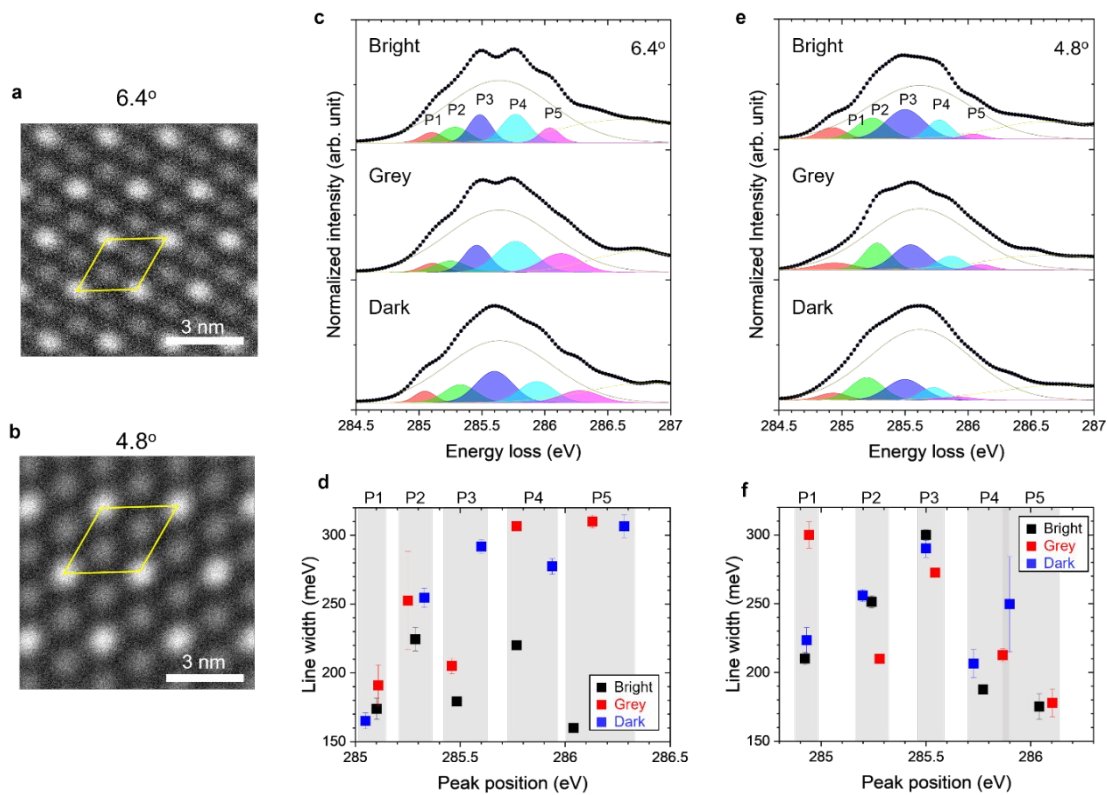

**Supplementary Figure 4. Local structure dependence of the fine structure of carbon K-edge for the twist angle of 6.4° and 4.8°. a**, **b**, STEM images of moiré superlattice of the twist angle of 6.4° and 4.8°. **c, e**, The fine structures of the π* region of the carbon K-edge corresponding to the bright, grey and dark regions with the results of the line shape analysis for the twist angle of 6.4° and 4.8°, respectively. The peak position and line width of peaks P1 to P5 in **c** and **e** are plotted in **d** and **f**.